# Control and ultrasonic actuation of a gas-liquid interface in a microfluidic chip

**Jie Xu and Daniel Attinger**
Laboratory for Microscale Transport Phenomena, Department of Mechanical Engineering, Columbia University, New York, NY 10027

E-mail: da2203@columbia.edu

**Abstract.** This article describes the design and manufacturing of a microfluidic chip, allowing for the actuation of a gas-liquid interface and of the neighboring fluid. A first way to control the interface motion is to apply a pressure difference across it. In this case, the efficiency of three different micro-geometries at anchoring the interface is compared. Also, the critical pressures needed to move the interface are measured and compared to theoretical result. A second way to control the interface motion is by ultrasonic excitation. When the excitation is weak, the interface exhibits traveling waves, which follow a dispersion equation. At stronger ultrasonic levels, standing waves appear on the interface, with frequencies that are half integer multiple of the excitation frequency. An associated microstreaming flow field observed in the vicinity of the interface is characterized. The meniscus and associated streaming flow have the potential to transport particles and mix reagents.

## 1. Introduction

Actuators based on the controlled motion of a gas-liquid interface have applications in microelectromechanical systems (MEMS) [1-8], with the ability to move fluid or particles. The generation of bubbles is usually performed using either thermal [1] or electrolytic [2] methods. Both methods induce phase change, and the bubble grows within milliseconds. During the fast thermodynamic transformation, the surrounding pressure, temperature and electrical field can experience drastic changes, with consequences for the liquid close to the bubble, for instance in biomedical applications. In this study, an alternative is explored, where a gas-liquid meniscus is slowly injected from a microchannel into a microchamber, using a syringe pump. Since the bubbles in most MEMS applications are attached to at least one wall [2, 4, 5], a meniscus might replace a bubble advantageously. For instance, the control of the meniscus position, volume and interfacial pressure is much simpler than for a bubble, and menisci can be moved in microchannels using capillary forces [9]. Also, such system does not experience temperature pulses. Regarding the actuation of the gas-liquid interface, two types of processes are used in MEMS devices: (1) a large displacement (usually greater than 10 μm) to directly push mechanical parts [3] or pump a liquid [5]; (2) high-frequency (on the order of 100 kHz [4]) oscillation of the interface with small amplitude to induce a steady microstreaming that mixes fluids or moves particles [7]. Using the first type of actuation process, Papavasiliou *et al.* [3] were able to displace a mechanical valve by about 100 μm; Deshmukh *et al.* [5] could drive a micropump at 0.5 μL/min (by periodic expansion and shrinking of a bubble, with the help of a passive microvalve), and Maxwell *et al.* [6] were able to capture and release particles by shrinking and expanding a microbubble in a cavity on the wall. Regarding the second type of actuation, Marmottant and Hilgenfeldt [7] observed a strong microstreaming field, in which particles can be transported. Also Kao *et al.* [4] showed that a micro-rotor can be driven at a speed as high as 625 rpm by the microstreamnig flow field induced by an oscillating bubble.

In this paper, we describe the manufacturing and characterization of microfluidic chip made of glass and cured Polydimethylsiloxane (PDMS). The potential of a gas-liquid meniscus to be used as an actuator in the chip is studied as follows. We first study the dynamics of large displacements of the meniscus (first type of actuation) by measuring the pressures needed to move the gas-liquid interface in a micro-geometry, and we classify the ability of different micro-features to anchor the interface at a desired location. Then, we study the response of the meniscus interface to high-frequency excitations. Besides a strong microstreaming flow in the vicinity of the interface, capillary traveling and standing waves are observed along the interface. We observe subharmonic capillary standing waves on the interface, which are analogous to Faraday waves [10]. Also, standing waves with superharmonic frequencies of the excitation frequency are observed. To the best of our knowledge, this is the first time that a superharmonic wave is found at a liquid-gas interface.

## 2. Chip design, fabrication and assembly

A typical microfluidic chip used in our study is shown in Figure 1: it involves a chamber (E) fed by a fork-like network of four channels A, B, C, D, with respective widths of 400, 1000, 400 and 100 micrometer. F is an interconnection needle and G is the clamp holding the chip. The height of each micro-channels is 50 µm. Channel D is used to inject air into the water-filled chamber E. The control of the gas-liquid interface position is enhanced by microgeometrical features at the junction of Channel D and the chamber, as shown in Figure 3 to 5. The microfluidic chip is manufactured using soft lithography [11] in the cleanroom of Columbia University, according to the following process. First, a 50µm thick SU-8 master (MicroChem) is cured on a silicon wafer with patterns transfered from a mask (CAD/Art Services Inc.). The chip is then manufactured from the master using PDMS Sylgard 184 Kit (Dow Corning). The PDMS chip is then covered with a PDMS cover plate and sandwiched between two glass slides hold by lateral clamps. This type of assembly ensures that all channel walls are made of PDMS and have the same surface properties. For some experiments involving high-frequency oscillations of the interface, a ring-shaped piezoelectric actuator is embedded in the PDMS cover plate, on top of the location shown by the letter *B* in Figure 1.

## 3. Experimental setup

The experimental setup is described in Figure 2, and involves three subsystems: a microfluidic platform, a piezoelectric actuation system and a visualization system. The microfluidic system involves the microfluidic chip described above and allows for the controlled filling of the chamber and subsequent controlled injection of an air plug in channel D, using a KDS 210 syringe pump (flow rate range from 0.001µl/hr to 86 ml/min). Visualization is performed with an Optem long-distance microscope in the plane perpendicular to the microfluidic chip, with a space resolution of about 1 micrometer. A strobe microscopy technique is used for visualizing the shape of the oscillating meniscus. A function generator (Agilent, 33120A) generates a square wave to both an amplifier (Krohn-Hite, 7600M) and a frequency divider. The amplifier controls the voltage applied to the piezoelectric transducer. The signal of the frequency divider is connected to a delay generator (BNC, 7010), which controls within one microsecond resolution the delay time between the diode illumination of the chamber and the actuation of the piezoelectric transducer. Therefore the diode frequency can be set to either the actuator frequency or half of its value, using the frequency divider.

The physical properties used in the experiments are described in Table 1. The surface tension, density and viscosity of water are from [12]. The contact angles are measured from the pictures using ScionImage.

**Table 1. Physical properties**

| Symbol | Physical property | Value |
|---|---|---|
| $\sigma$ | Surface tension of water | 72.0 mN/m @ 25 °C |
| | | 69.6 mN/m @ 40 °C |
| $\theta$ | Contact angle of water on PDMS | 70° for receding |
| | | 110° for advancing |
| $\rho$ | Density of water | 0.998 g/cm$^3$ |

## 4. Characterization of meniscus dynamics

This section investigates the meniscus motion during the injection phase, as well as the response of the meniscus to ultrasonic excitation.

*4.1 Ability of three microgeometries to trap a contact line*

Repeatability is important for designing a MEMS actuator, and typical free-floating bubbles are hard to manipulate and can dissolve [13, 14]. In this work, we investigated the efficiency of microgeometrical features to trap the meniscus contact line during the air injection process. Three types of features named *pits*, *peaks* and *overhangs* are shown in the respective Figure 3, 4 and 5, and are compared. The sequential frames in Figure 3-5 show the time evolution of the interaction between the moving meniscus and the geometrical feature. From Figure 3, we observe that *pits* are not able to trap the meniscus contact line, while the *peaks* and *overhangs* in Figure 4 and 5 pin the meniscus contact line efficiently. A measure of the 'pinning efficiency' is the range of meniscus visible area for which the contact line is pinned. Using image analysis of a movie taken during meniscus growth, we define $S_1$ as the visible area when the contact line starts to be pinned and $S_2$ as the area when the contact lines moves again. The measured value $\Delta S = S_2 - S_1$ is indicated in Figures 3-5. The large values of $\Delta S$ *in* the *peaks* and *overhangs* (0.011 and 0.021 mm$^2$ respectively) confirm that their ability at trapping the contact line is better than the *pit* geometry, where $\Delta S$ *is* negligible.

The better ability of the *overhangs* and *peaks* at trapping a contact line is explained as follows. For the *pits* design in Figure 6(a), assuming the meniscus keeps a perfect circular shape during its growth and neglecting the influence of the top and bottom walls, when the contact line reaches Location 1, the rest of the interface is very close to the other side of the pit. Therefore, a slight increase of the meniscus volume can induce a contact between the meniscus and the other side of the pit, which implies a wetting angle value close to zero and a fast receding of the contact line: the meniscus passes over the pit. This explains why $\Delta S$ for *pits* is nearly zero in Figure 3. For the *peaks* and *overhangs* design, when the contact line reaches Location 1, as shown in Figure 6 (b) and (c), the geometry starts to trap the contact line. At this time the contact angle $\theta_1$ is defined to be the angle between the contact line and the holding surface, with the receding contact angle value, namely, 70°. The contact line then rotates until it reaches Location 2, where the contact angle $\theta_2$ is defined to be the angle between the contact line and the releasing surface, which also equals the receding contact angle value. Further growth of the meniscus will move the ends of the contact line.

Therefore, for *peaks* the contact line is pinned between Location 1 and 2, and the $\Delta S$ is calculated to be 0.042 mm$^2$. In Figure 6(c), we can see that for the *overhangs*, $\theta_2$ will never reach 70° even if the meniscus grows to infinity. Therefore, the theoretical value of $\Delta S$ for *overhangs* should be very large. This explains why that design is best at stopping the progression of the meniscus, with a value $\Delta S$ of 0.021 mm$^2$. While this theoretical analysis does not allow to quantitatively predict the ability of a geometry to trap a contact line, it is useful at classifying the relative ability of various designs, and can be used for future design. Also, our experiments showed that the *peaks* design is the best at holding a meniscus in a circular shape while the *overhangs* design gives the largest stable area. Both features are useful for bubble-based actuators.

*4.2 Pressure measurement across the meniscus*
Another way to study the meniscus stability is to measure the pressure amplitude that can be applied across the meniscus while keeping the contact line pinned on the micro-features of Figures 3-5. The pressure is measured using a differential pressure sensor (Honeywell 143PC03D with pressure range of ±17 kPa): one sensor port is connected to the outlet of Syringe 1 and the other port is open to the atmosphere, as shown in Figure 2. The pressure resolution uncertainty is 0.2 Pa, which is due to the voltage measurement. The measurement is performed on a chamber with the *peaks* microgeometry of Figure 4: starting from a meniscus at equilibrium, we either increase or decrease the pressure inside the meniscus by slowly operating the syringe pump, at a typical rate of 0.1 mL/min. The pressure at the instant where the meniscus starts to grow is called *upper critical pressure*. The pressure at the instant where the meniscus starts to shrink is called *lower critical pressure*.

These critical pressures are plotted in Figure 7 as a function of the meniscus radius. Both lower and upper critical pressure decrease with the meniscus radius and their difference is constant, at a value of about 2000 Pa. This behavior can be explained as follows. Across the air-water meniscus, the pressure difference is expressed by the Laplace equation [15]:

$$\Delta P = \sigma \left( \frac{1}{R_h} + \frac{1}{R_w} \right) \quad (1)$$

where $R_w$ and $R_h$ are radii of curvature in the respective parallel and perpendicular plane to the microfluidic chip. Experimentally we observe that when the pressure difference $\Delta P$ is modified, $R_h$ changes first while $R_w$ stays constant. The modification of $R_h$ occurs by a variation of the wetting angle between its receding value 70° and advancing value 110°. $R_h$ can be expressed as

$$R_h = \frac{h}{2\cos\theta} \quad (2)$$

where $\theta$ is the advancing or receding wetting angle, with chamber thickness $h=50$ μm. Once the pressure modification is too large to be compensated by the sole variation of $R_h$, $R_w$ changes and induces a visible motion of the meniscus interface in the visualization plane (see e.g. Figure 4). It is worth noting that there is a position where $R_w$ has a minimum value, which equals half of the width of the microchannel, namely, 50 μm. This implies that once $R_w$ starts changing, the motion of the meniscus becomes unstable, either rapidly expanding or rapidly shrinking, in virtue of (1), similarly to the dynamic of an air bubble popping out of an underwater needle. The above theory predicts a range of $\Delta P$ where the visible shape of the meniscus does not change (i.e. $R_w$ is constant), while $R_h$ varies within the bounds of the wetting angle $\theta$ (see (2) and Table 1). Using the surface tension of water at room temperature, this theoretical range of $\Delta P$ is plotted as two solid lines as a function of $R_h$ in Figure 7 and agrees relatively well with our measurements. The uncertainty in Figure 7 is determined by

$$u = \left(u_0^2 + u_1^2 + u_2^2\right)^{1/2} = 136 Pa \qquad (3)$$

where $u_0$ is the pressure resolution uncertainty (0.2Pa), $u_1$ is the linearity error of the sensor (129 Pa) and $u_2$ is the repeatability and hysteresis error of the sensor (43 Pa). Both the use of micro-geometries to trap a meniscus and the measurement of pressure hysteresis that keep the meniscus stable can benefit the design of MEMS systems where a gas-liquid interface has to be controlled in position.

*4.3 Response to ultrasonic excitation*
We also studied the meniscus response to high-frequency pressure oscillations. To induce these oscillations, we cure in the PDMS cover layer a piezoelectric transducer (ring-shaped PZT with 2mm thickness). The transducer drives pressure waves through the water in the microchamber that induce various kind of surface waves on the water/air interface, as well as a streaming flow in the liquid close to the meniscus, phenomena which are described as follows.

*4.3.1 Traveling wave.* For a discrete set of excitation frequencies, we observed waves traveling on the meniscus surface with the same frequency as the excitation. Figure 8 shows a sequence taken at 150 kHz of excitation frequency, with a delay of 2 μs between each frame. Crests A and B are moving towards the center of symmetry of the meniscus. This phenomenon also occurs when the meniscus is retracted and assumes a flat shape in the chip plane, as shown in Figure 9: this configuration simplifies the measurement of the wavelength, as well as the analysis of the phenomenon. Right to the individual frames in Figure 9 are the values of the observed wavelengths at the given oscillation frequencies. Since the wavelengths are small ($\lambda \langle\langle 2\pi\kappa^{-1}$, where $\kappa^{-1}$ is the water-air capillary length, with $2\pi\kappa^{-1}$ typically on the order of a centimeter), gravity can be neglected in the theoretical analysis: these waves are capillary waves, caused by inertia and surface tension. We observe that the wavelength decreases as the excitation frequency increases, and this measurement is plotted in Figure 11. The wavelength-frequency relationship can be described in the classical framework of surface waves analysis [16], neglecting gravity. Figure 10 defines the nomenclature and reference frame with the *x-z* plane parallel to the microfluidic chip. In a pure 2-D case in the *x-z* plane, a dispersion relation exists that relates the wavelength to the wave frequency [16]:

$$\lambda = \left(\frac{2\pi\sigma}{f^2\rho}\right)^{\frac{1}{3}} \qquad (4)$$

This relation is plotted in Figure 11, for a water surface tension at the temperature of 40°C, corresponding to the measured water temperature due to the resistive dissipation in the piezoelectric. Both theory and experiment show the trend that wavelength decreases as the frequency increases, with the predicted wavelength approximately 5% larger than the observed wavelength. This slight discrepancy might come from the fact that in deriving (4), only the capillary restoring force due to the curvature in the plane of the chip (*x-z* plane) is considered. In the microfluidic chip however, the capillary restoring force also depends on the curvature in the perpendicular plane (*x-y* plane), see (1). This curvature is caused by the wave amplitude and $z_0$ as defined in Figure 10. These two contributions explain why the visible interface is blurred in Figure 8. The total contribution of $z_0$ (about 2 μm) and wave amplitude (about 2 μm) to the curvature in the perpendicular plane can be expressed as

$$\frac{1}{R_h} = -\frac{2(\eta - z_0)}{(\eta - z_0)^2 - h^2/4} \qquad (5)$$

while the other curvature, in the plane of the chip is

$$\frac{1}{R_w} = \frac{-\partial^2 \eta / \partial x^2}{[1+(\partial \eta / \partial x)^2]^{3/2}} \approx -\frac{\partial^2 \eta}{\partial x^2} \tag{6}$$

By assuming a sine-shape wave with wavelength values as reported in Figure 11, $R_w$ is on the order of 10 μm, and $R_h$ is much larger on the order of 80 μm, which justifies the 2D analysis presented here. Also, the good agreement in Figure 11 between the experimental data and the 2-D theory points to the fact that the flow field in our system is mostly 2-D, which will greatly simplify future developments of similar chips.

*4.3.2 Standing wave.* Increasing the actuation intensity –while keeping the excitation frequency $f_e$ constant at 150kHz- generates a standing wave, which is described by the successive frames in Figure 12. In that case, the standing wave has a larger amplitude (around 5 μm) than the traveling wave (around 2 μm, see Figure 8). Also, the standing wave is a subharmonic of the excitation frequency, with the wave frequency $f_1 = 1/2\ f_e$. Similarly to the experiments with the traveling wave, a standing wave is also observed for a flat meniscus (Figure 13). Typical experimental conditions at which these standing waves occur are listed in Table 2, with the amplifier voltage as an indicative measurement of the needed excitation power. Table 2 shows that standing waves can be either subharmonic, harmonic or superharmonic of the excitation frequency, in contrast with traveling waves, which are always observed to be harmonic.

**Table 2. Conditions at the transition from traveling to standing waves microstreaming**

| $f_e$ (kHz) | $f_1/f_e$ | $f_1$ (kHz) | Smallest voltage needed to induce standing wave (V) |
|---|---|---|---|
| 175 | ½ | 87.5 | 98.6 |
| 169 | ½ | 84.5 | 147 |
| 165 | ½ | 82.5 | 121.4 |
| 154 | ½ | 77 | 77.8 |
| 150 | ½ | 75 | 29.9 |
| 145 | ½ | 72.5 | 36.3 |
| 139 | ½ | 69.5 | 60.3 |
| 127 | ½ | 63.5 | 77.4 |
| 76 | 1 | 76 | 182.3 |
| 75 | 1 | 75 | 182.8 |
| 74 | 1 | 74 | 182.3 |
| 51 | 3/2 | 76.5 | 115.3 |
| 50 | 3/2 | 75 | 107.7 |
| 49 | 3/2 | 73.5 | 95.1 |

An analogous observation of subharmonic waves is the Faraday instability, where gravity and inertia drive waves at a frequency $f_1=f_e/2$ [10]. In our experiment, the ratios of observed wave frequencies $f_1$ to excitation frequencies $f_e$, given in Table 2, were all multiple of $f_e/2$, and could therefore be observed with the frequency divider. For standing waves, the allowable wavelengths for a condition of no flow through the walls are

$$\lambda = \frac{L}{n+1} \tag{7}$$

where $L$ is the meniscus length and $n=0,1,-,n$ [17]. Spontaneously, the system chooses $n=1$, with $\lambda=50$ μm on a 100 μm meniscus such as in figure 13. Subharmonic responses at gas-liquid interfaces have often been observed [10, 18-21]. However this is to the best of our knowledge the first time that capillary waves with superharmonic frequencies of the excitation frequency are observed, although this has been theoretically demonstrated by Eisenmenger [22] for the related case of gravity-driven Faraday waves.

*4.3.3 Microstreaming.* The occurrence of a microstreaming flow in the vicinity of the meniscus has been observed with microdroplets of liquid PDMS (Alfa Aesar, M.W. 770) dispersed in water. In Figure 14, two vortices are found on the left and right of the meniscus and the flow direction is indicated, in a pattern that is consistent with observations by Marmottant, *et al*. in a 3D case [23].

An estimation of the flow speed in a 20 μm circle with center at x = 120 μm, z = 50 μm denoted as region A in Figure 14 was performed by dispersing 1.0 μm Polymer Microspheres (Duke Scientific) in the flow and visualizing their trajectories with a camera at up to 280 frames per second. Figure 15 shows the flow speed as a function of excitation frequency, a fixed voltage of 20V being applied to the piezoelectric transducer. The velocities are determined from the displacement *l* of a microsphere in a time *dt*. The root-sum-square uncertainty on the velocity magnitude is estimated assuming that the microsphere follow the flow perfectly, and assuming conservative uncertainties *Δl* of 5 μm for the displacement, and *Δt* the half period between each visualization frame:

$$\Delta v = \left( \left( \frac{\Delta l}{dt} \right)^2 + \left( \frac{l \Delta t}{dt^2} \right)^2 \right)^{1/2} \quad (8)$$

Figure 15 shows that the flow speed reaches a maximum value close to 1mm/s for an excitation at 140 kHz. This frequency is close to the frequency where a standing wave is induced on the meniscus (see Table 2) for the smallest excitation intensity, an indication that maximizing the streaming is related to exciting the meniscus at its natural frequency. In Figure 16, the flow speed is plotted for a fixed excitation frequency (100 kHz), for increasing values of the excitation intensity. It appears that the streaming speed grows quadratically with the ultrasonic intensity. This quadratic relation is characteristic of the second-order streaming flow [24]. Note that microstreaming was observed in both traveling and standing wave regime. However, in the standing wave case, the meniscus position was less stable than in the traveling wave case. Clearly this microstreaming flow has a potential to attract particles or mix liquid streams.

5. Conclusion

We have described the manufacturing of microfluidic chips for bubble-based actuators. First, we quantitatively identified the efficiency of three micro-geometries at trapping the contact line of the gas-liquid meniscus. Theoretically, we found that each meniscus radius corresponds to a range of pressure that holds the meniscus stable, because of the hysteresis of the wetting angle, and this finding is confirmed by pressure measurements. In a second part of our study we investigate the meniscus response to ultrasonic excitation. When the excitation is weak, traveling waves are found with a wavelength that decreases with increasing excitation frequency, which is well predicted by capillary wave theory. When the excitation becomes stronger, standing waves at the meniscus interface with different ratios (1/2, 1, 3/2) of wave frequency to excitation frequency are observed. The magnitude of the associated streaming flow due to the meniscus oscillations is studied in relation to the excitation frequency and intensity. The meniscus and associated streaming flow have the potential to transport particles and mix reagents in a microfluidic chip.

**Acknowledgement**: We thank the US National Science Foundation for their support through the CAREER grant 0449269.

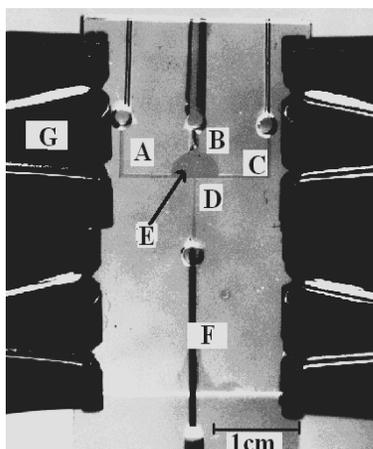

**Figure 1: A typical microfluidic chip involving a chamber fed by a fork-like network of four channels.**

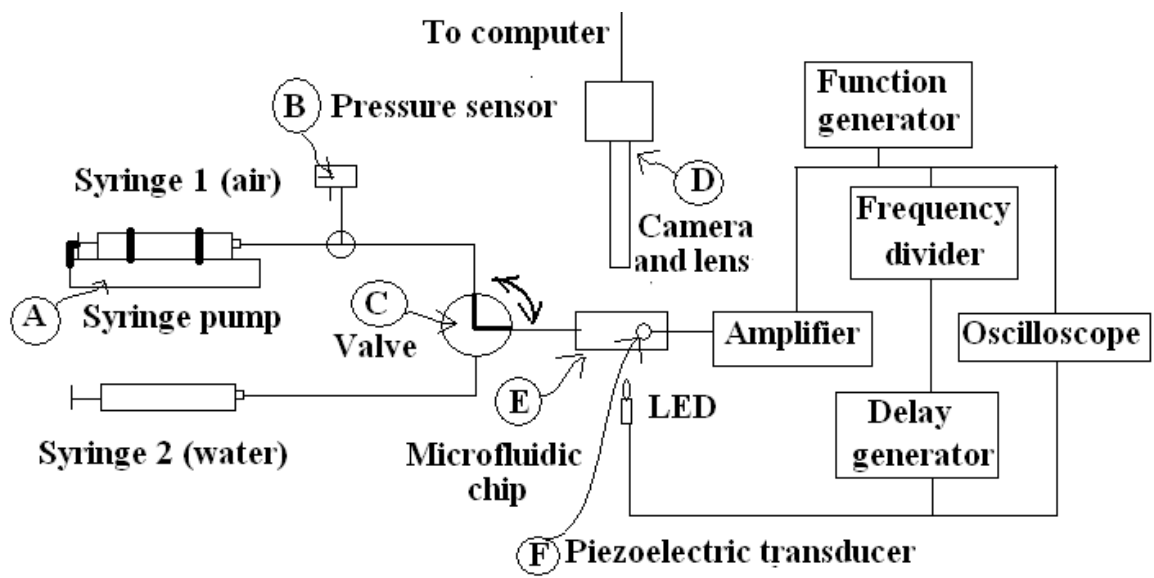

Figure 2: The experimental setup involves: a microfluidic platform, a piezoelectric actuation system and a visualization system, During the experiment, we connect the valve C to syringe 2, so that water is infused into the system. Then, the valve is shifted to syringe 1 to add a controlled flow rate of air into the system and forms a meniscus.

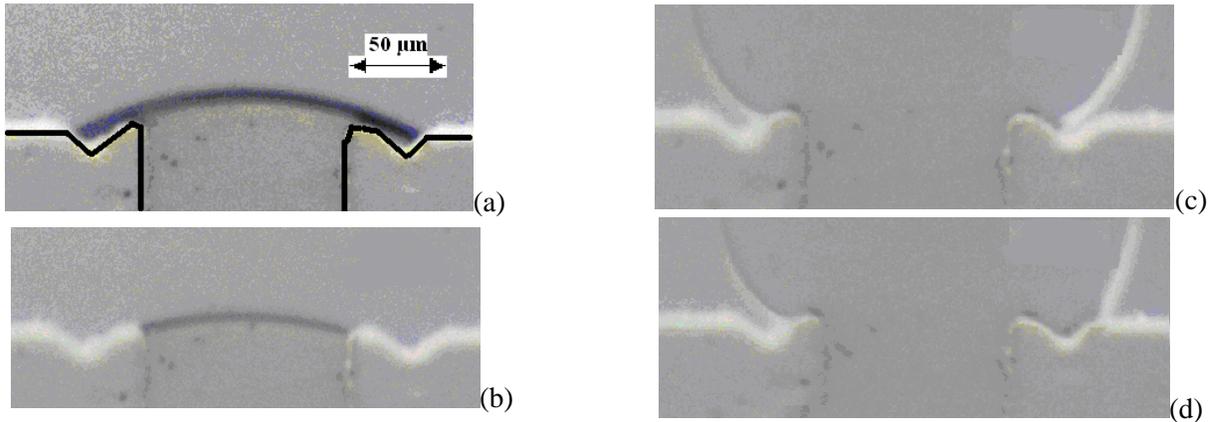

Figure 3: Microgeometry with pits at the entry of channel D into chamber E. Sequences (a-b) and (c-d) show that the meniscus contact line is not pinned by the pits, when the meniscus shrinks and grows respectively. ΔS in this case is 0 mm².

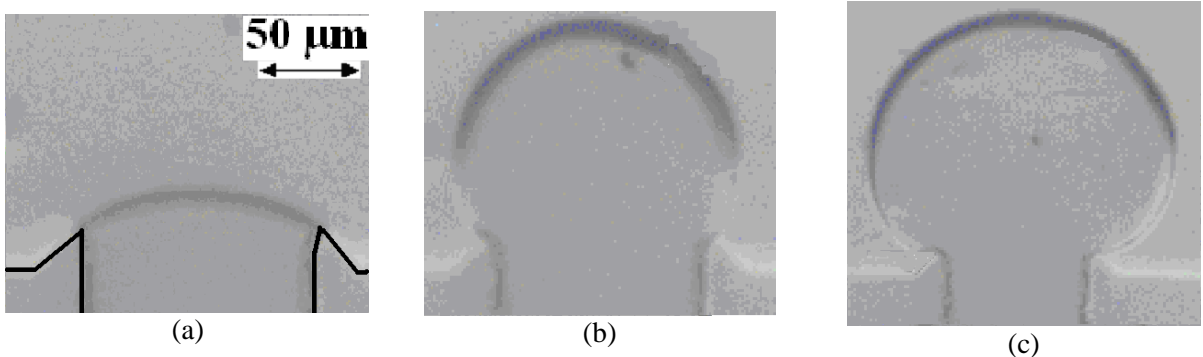

Figure 4: Microgeometry with peaks at the chamber entry. Sequence (a-c) show how the meniscus contact line is pinned at the peaks. ΔS in this case is 0.011 mm².

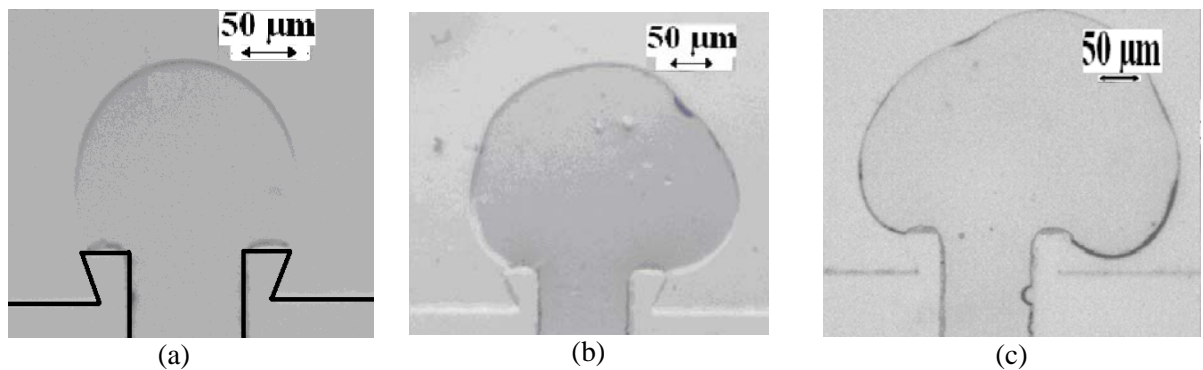

Figure 5: Microgeometry with overhang. Sequence (a-c) shows and excellent pinning of the contact line. ΔS in this case is 0.021 mm².

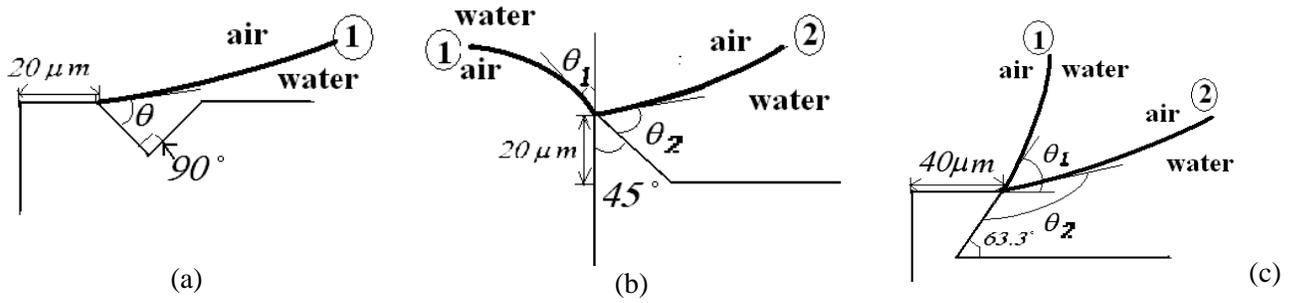

(a)                         (b)                        (c)

**Figure 6: Theoretical analysis for the contact line stable range during bubble growth.**

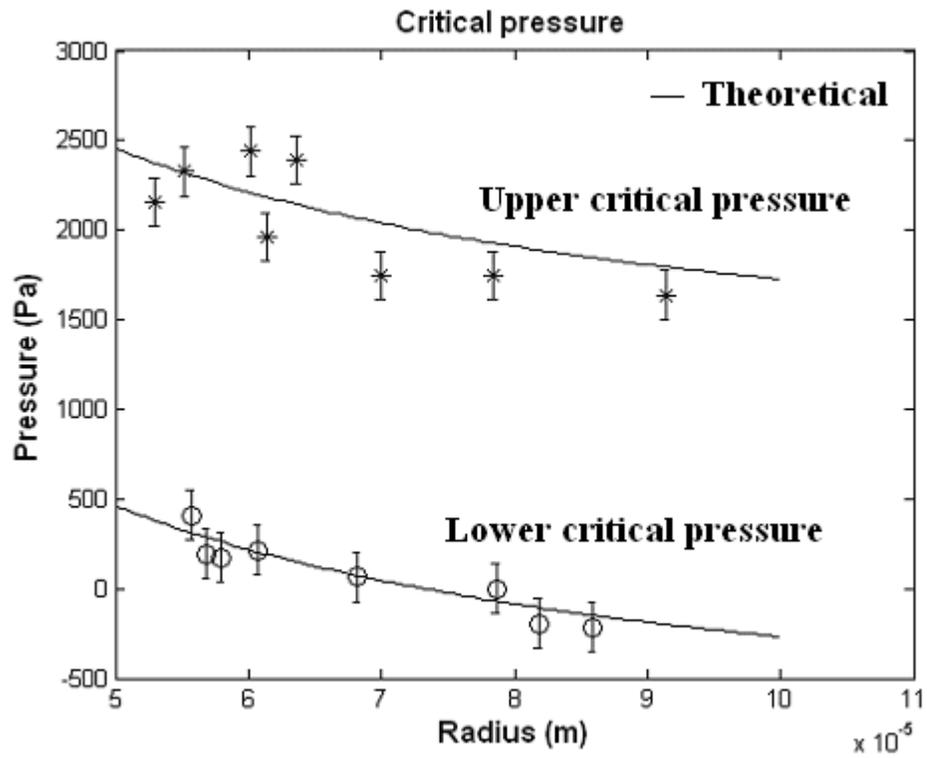

**Figure 7: Study of critical pressures across the meniscus.**

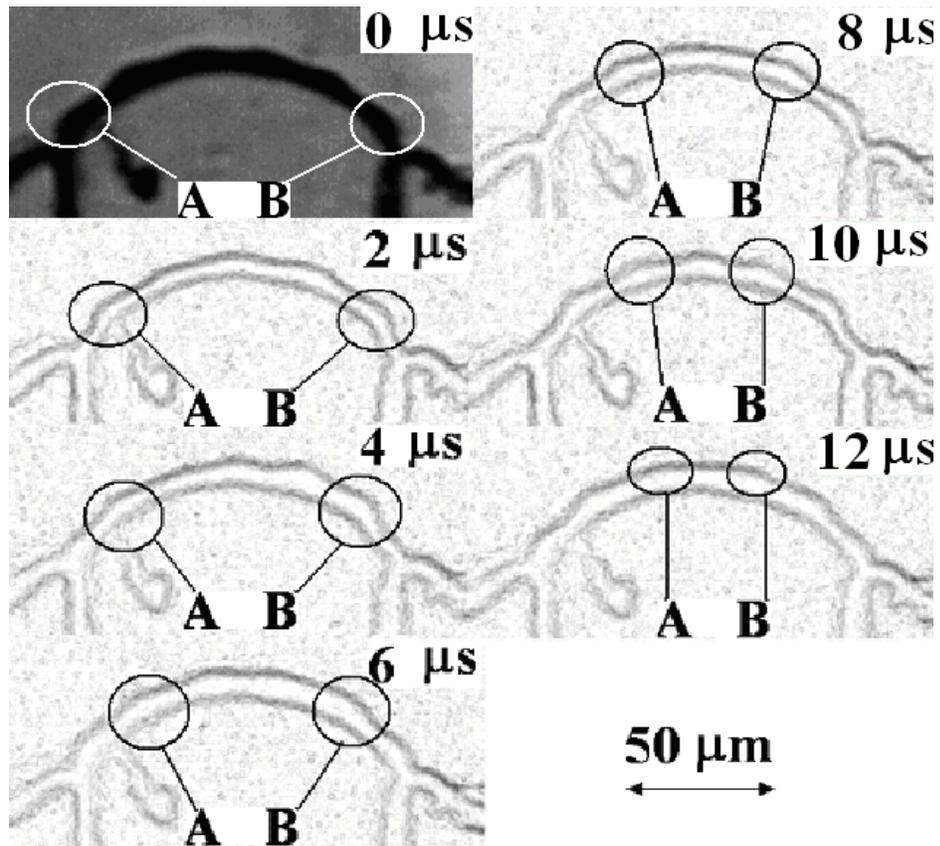

**Figure 8: Traveling waves on meniscus at 150 kHz. The pictures at t>0 have artificially enhanced contrast using an edge detection algorithm from Adobe Photoshop.**

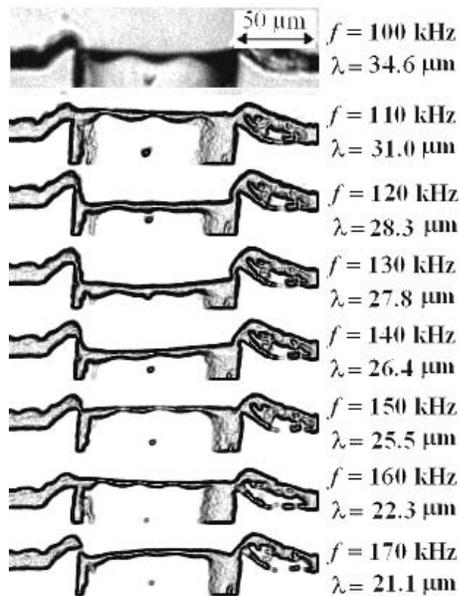

**Figure 9: Traveling waves at different frequencies. The pictures for *f*>100 kHz have artificially enhanced contrast using an edge detection algorithm from Adobe Photoshop.**

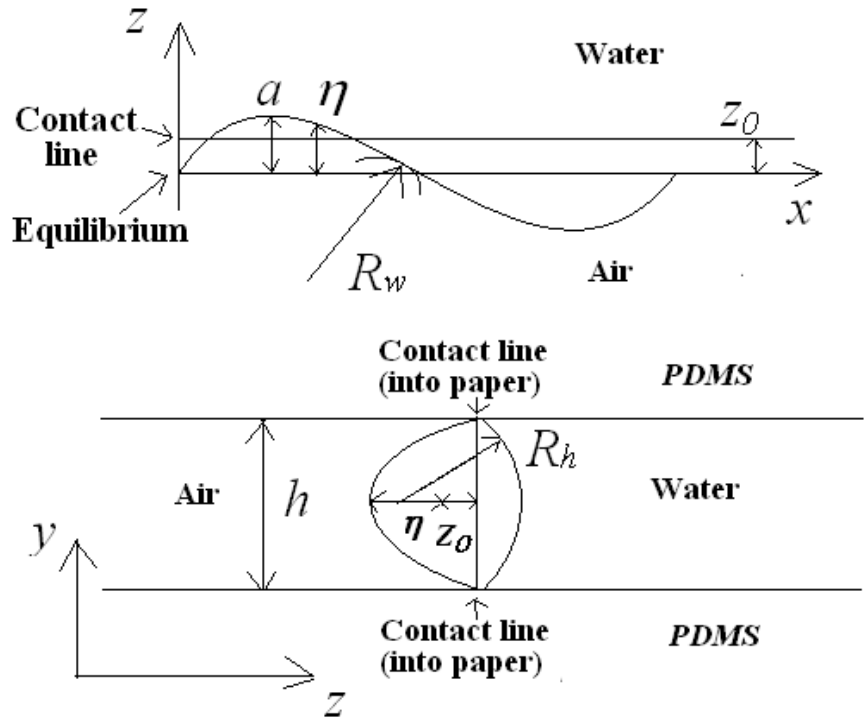

**Figure 10: 3D shape of the meniscus oscillation, where $z_0$ is the distance between contact line and the $z$-location corresponding to $\eta = 0$.**

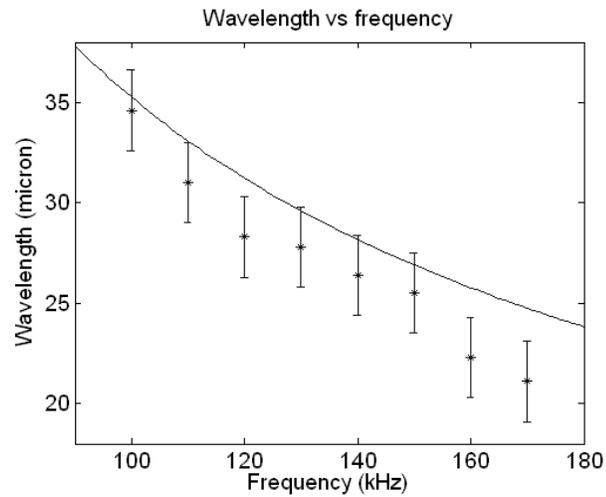

**Figure 11: Wavelength vs frequency**

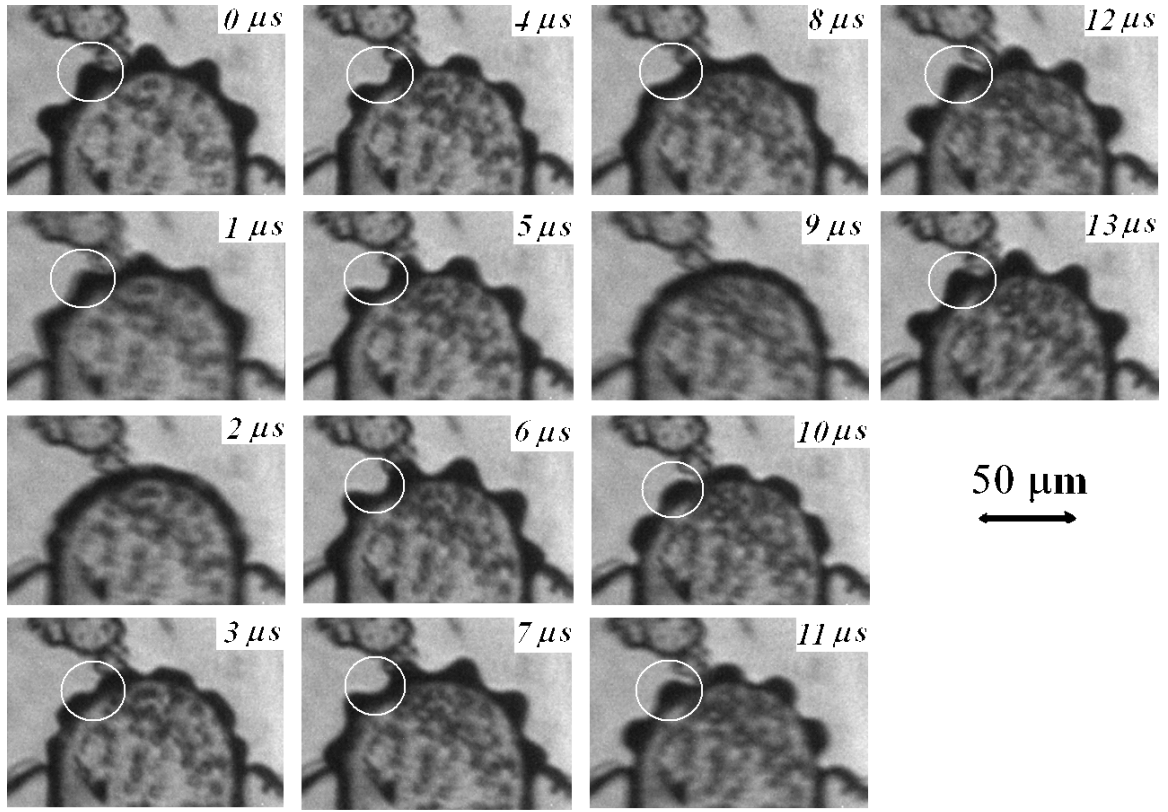

**Figure 12: Standing wave on meniscus interface at 150 kHz of excitation, the standing wave is oscillating at 75 kHz. Note that there are drops generated in the air, which are found more often in the standing wave case than the traveling wave case.**

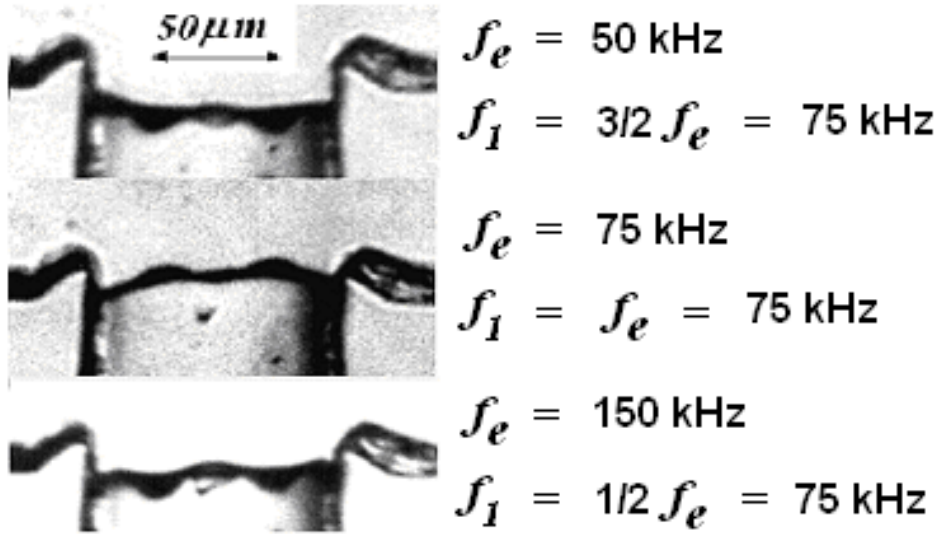

**Figure 13: Standing waves at different frequencies.**

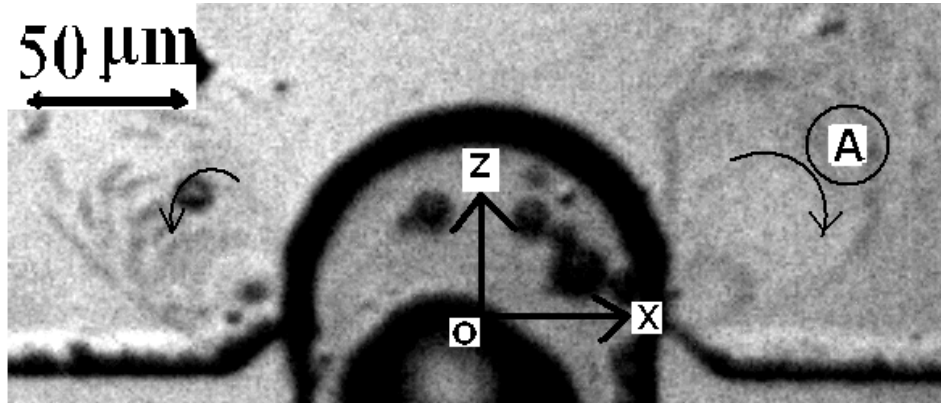

Figure 14: Two microstreaming vortices visible at the left and right of the meniscus vicinity. The excitation frequency is 110V and the voltage is 90 V. Arrows indicate the direction of rotation.

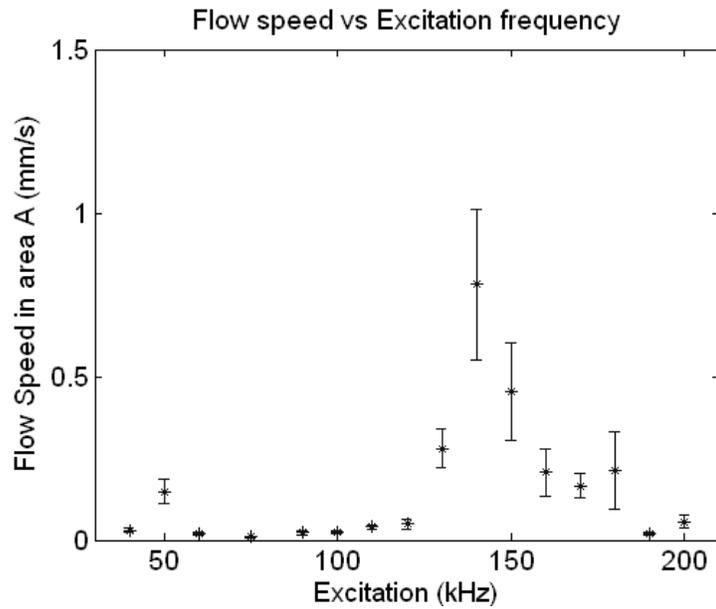

Figure 15: Flow speed in area A (Figure 16) vs excitation frequencies.

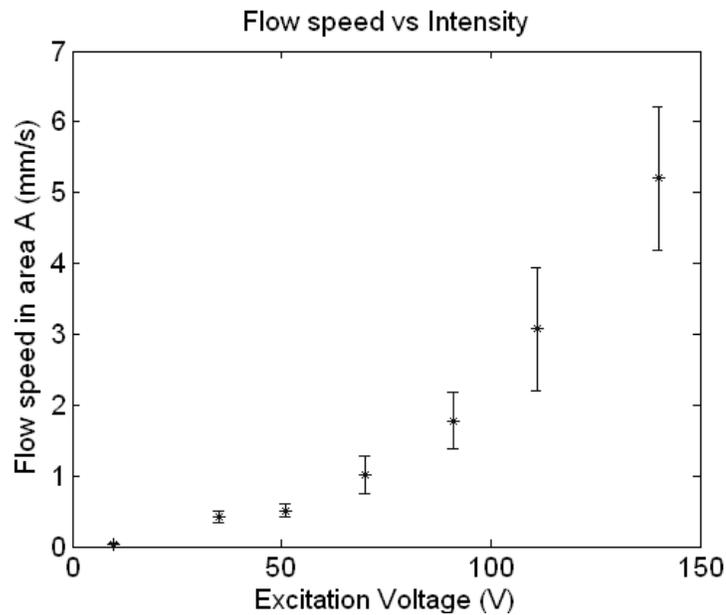

Figure 16: Flow speed in area A of Figure 16 vs intensity of the piezoelectric excitation. The excitation frequency is 100 kHz